\documentstyle[12pt,aaspp4]{article}

\lefthead{Briggs et al.}
\righthead{BATSE Observations of GRB Isotropy}

\begin{document}

\thispagestyle{empty}

\vspace*{0.1in}
\title{BATSE Observations of the Large-Scale \\ Isotropy of Gamma-Ray Bursts}
\vspace{0.35in}

\author{Michael S. Briggs, William S. Paciesas,
and Geoffrey N. Pendleton}
\affil{Department of Physics, University of Alabama in Huntsville,
Huntsville, AL 35899}

\author{Charles A. Meegan, Gerald J. Fishman,  John M. Horack, \\
and Martin N. Brock}
\affil{NASA/Marshall Space Flight Center, Huntsville, AL 35812}

\author{Chryssa Kouveliotou}
\affil{Universities Space Research Association \\
NASA/Marshall Space Flight Center, Huntsville, AL 35812}

\author{Dieter H. Hartmann}
\affil{Department of Physics and Astronomy \\
Clemson University, Clemson, SC 29634}

\and

\author{Jon Hakkila}
\affil{Department of Physics and Astronomy \\
Mankato State University, Mankato, MN 56002}

\vspace{0.35in}
\begin{center}
To appear in {\it The Astrophysical Journal} on 1996 March 1. \\
\copyright \hspace{0.5mm} 1995 by The American Astronomical Society.
All Rights Reserved.
\end{center}

\hfill

\clearpage
\thispagestyle{empty}

\vspace*{0.25in}
\begin{abstract}

We use dipole and quadrupole statistics to test the large-scale isotropy
of the first 1005 gamma-ray bursts observed by the Burst and Transient
Source Experiment (BATSE).
In addition to the entire sample of 1005 gamma-ray bursts, many subsets are
examined.
We use a variety of dipole and quadrupole statistics to
search for Galactic and other predicted anisotropies and for
anisotropies in a coordinate-system independent manner.
We find the gamma-ray burst
locations to be consistent with isotropy, e.g.,
for the total sample the observed Galactic dipole moment
$\langle \cos \theta \rangle$ differs from the value predicted for isotropy
by $0.9 \sigma$ and the observed Galactic quadrupole moment
$\langle \sin^2 b - \frac{1}{3} \rangle$ by $0.3 \sigma$.
We estimate for various models the anisotropies that could have been
detected.
If one-half of the locations were within $86^\circ$ of the Galactic
center, or within $28^\circ$ of the Galactic plane, the
ensuing dipole or quadrupole moment
would have typically been detected at the 99\% confidence level.
We compare the observations with
the dipole and quadrupole moments of various Galactic models.
Several Galactic gamma-ray bursts models have moments within $2\sigma$ of the
observations;
most of the Galactic models proposed to date are no longer in acceptable
agreement with the data.
Although a spherical Dark Matter Halo distribution could be
consistent with the data, the required core radius is larger than the
core radius of the Dark Matter Halo used to explain the Galaxy's rotation
curve.
Gamma-ray bursts are much more isotropic than any observed Galactic
population, strongly favoring but not requiring an origin at cosmological
distances.

\end{abstract}


\hfill
\clearpage

\thispagestyle{empty}
\setcounter{page}{1}

\section {Introduction}

The presence of a large-scale pattern in the locations of gamma-ray
bursts (GRBs) would be a major clue to their origin.
However, the gamma-ray burst locations determined by the Burst and
Transient Source Experiment (BATSE) on the Compton Gamma-Ray Observatory (CGRO)
remain consistent
with large-scale isotropy.
This indicates either that the GRBs are isotropic or that
their anisotropy is too small to have yet been
detected.
The upper-limits on the magnitudes of anisotropies strongly
constrain Galactic models.

One possible reason for not detecting an anisotropy is that
we are observing gamma-ray bursts to a depth
much less than the length scale of their distribution.
This was thought to be the case prior to BATSE, when it was generally believed
that gamma-ray bursts originate from a disk population of neutron stars
(Hurley 1986; Higdon \& Lingenfelter 1990; Harding 1994).
However, this possibility
is excluded by BATSE's concurrent observation that gamma-ray bursts are
inhomogeneous, i.e., there is a deficiency of faint GRBs compared to
a uniform density in Euclidean space.
The inhomogeneity of the GRBs is demonstrated
by $\langle V/V_{\rm max} \rangle < 0.5$
and by the deviation of the
$\log{(N>P)}-\log{P}$ distribution from a $-$\slantfrac{3}{2} slope
power law for the fainter
GRBs (Meegan et al. 1992a;
Fishman et al. 1994;
Meegan 1994a; Horack \& Emslie 1994).
The effect is not small:
the number of GRBs with peak flux $P$ = 0.8
photons s$^{-1}$ cm$^{-2}$ between 50 and 300 keV on the 256 ms timescale
is deficient by a factor of about six relative
to the extrapolation of the
$-$\slantfrac{3}{2} slope power law characteristic of the homogeneous
GRBs brighter than 10 photons s$^{-1}$ cm$^{-2}$
(Pendleton et al. 1995).
Assuming a galactic origin, the observed inhomogeneity
means that instead of viewing to less than a length scale, BATSE is viewing
many length scales and sees past the ``edge'' of the distribution.

Another possible explanation for not detecting an anisotropy, even though
we are seeing past the edge of the distribution, is that our offset from
the center of the distribution is very small.
A local version of this possibility is that
gamma-ray bursts originate from a cometary cloud around the Sun
(Bickert \& Greiner 1993; Katz 1993; White 1993; Luchkov 1994).
In this hypothesis,
the Earth-Sun distance is too small relative to the diameter of the comet cloud
to cause a detectable anisotropy and the edge of the comet distribution causes
the inhomogeneity.
A distant version of this possibility is that gamma-ray bursts originate from
a Galactic halo, in which case the length scale of the halo must be
large compared to the distance from the Sun
to the center of the Galaxy
(Fishman 1979;
Jennings 1982;
Shklovskii \& Mitrofanov 1985;
Atteia \& Hurley 1986).
Finally, gamma-ray bursts might be isotropic because
they originate at cosmological distances
(Prilutski \& Usov 1975;
Usov \& Chibisov 1975;
van den Bergh 1983;
Paczy\'{n}ski 1986).
In this hypothesis,
the observed inhomogeneity is due either to the non-Euclidean
geometry of the Universe, to source evolution, or to both.
If GRBs come from $z \lesssim 1$, too few have been observed to detect
the dipole moment due to our motion with respect to distant matter
(Brainerd 1995).

Arguably, the most crucial question about gamma-ray bursts is their distance.
While discovering or placing stringent limits on large-scale
anisotropies promises to disclose the distance scale, discovering gamma-ray
burst repeaters would not, since there are cosmological models allowing
or predicting repetition
(McBreen \& Metcalfe 1988;
Dermer \& Schlickeiser 1994;
McBreen, Plunkett \& Metcalfe 1993;
Brainerd 1994).    The discovery of GRB repetition would, however,
disprove the most popular model for a cosmological origin, the mergers
of compact stars.

The BATSE team  has described its
findings that gamma-ray bursts are isotropic and inhomogeneous
in previous papers and circulars
(Meegan et al. 1991; Meegan et al. 1992a; Meegan et al. 1992b;
Briggs et al. 1993a; Briggs et al. 1993b; Meegan et al. 1993a;
Horack et al. 1993; Fishman et al. 1994; Horack et al. 1994;
Meegan et al. 1994a; Briggs et al. 1994).
Many authors have emphasized that the isotropic and inhomogeneous
GRB spatial distribution observed by BATSE is
natural if gamma-ray bursts
originate at cosmological distances
(Paczy\'{n}ski 1991a;
Dermer 1992;
Mao \& Paczy\'{n}ski 1992a;
Piran 1992;
Fenimore et al. 1993;
Wickramasinghe et al. 1993;
Woods \& Loeb 1994).
Paczy\'{n}ski \& Xu (1994) state that
``A conservative conclusion is that the sources are at cosmological
distances.''
Other authors have emphasized the possibility of a Galactic origin with
an anisotropic pattern too small to have been detected with BATSE, such as
a very extended halo distribution or a combination of source
distributions
(Brainerd 1992;
Eichler \& Silk 1992;
Hartmann 1992;
Li \& Dermer 1992;
Lingenfelter \& Higdon 1992;
Atteia \& Dezalay 1993;
Fabian \& Podsiadlowski 1993;
Liang \& Li 1993;
Smith \& Lamb 1993;
Hartmann et al. 1994a;
Higdon \& Lingenfelter 1994;
Li, Duncan \& Thompson 1994;
Lyne \& Lorimer 1994;
Podsiadlowski, Rees \& Ruderman 1995).
These authors have generally argued that GRB spectral and temporal
properties make a Galactic origin likely.
Quashnock \& Lamb (1993a) identified a subset
of the first BATSE (1B) catalog (Fishman et al. 1994)
which has a concentration
towards the Galactic plane and center, a pattern
which they interpreted as demonstrating that all gamma-ray bursts originate in
the spiral arms of the Milky Way.

In this paper we test the isotropy of the locations of a larger sample,
the first 1005 gamma-ray bursts detected by BATSE.
Isotropy means that the probability of a GRB occurring in a given region
of the sky is solely proportional to the solid angle of that region.
While isotropy requires the absence of patterns on all angular scales, herein
we emphasize the large angular scale properties of gamma-ray bursts
by testing isotropy with dipole and quadrupole tests.
We will analyze many subsets, including ones generated by the criteria
of Quashnock \& Lamb (1993a).   We show that the locations are
consistent with isotropy, derive upper limits on the parameters
of several anisotropic models, and compare the observed moments with those
of Galactic models.

The most plausible anisotropy to which dipole and quadrupole tests are
insensitive is small-scale clustering,
including GRB repetition, which averages to
isotropy on large angular scales.
Tests sensitive to such anisotropies include those based upon the
nearest neighbor distribution and the two-point angular correlation function.
Meegan et al. (1995), Hartmann et al. (1994b), and Brainerd et al. (1995)
apply these tests and others to the data of the
second BATSE catalog (2B) (Meegan et al. 1994b) and find no evidence for
small-scale anisotropies in BATSE's GRB locations.
While with the 2B catalog these authors
do not confirm the evidence for GRB repetition found by
Quashnock \& Lamb (1993b) and by Wang \& Lingenfelter (1993; 1995) in
their analyses of the 1B catalog, they do not
completely exclude burst repetition:
the 99\% confidence upper-limit placed on the repeater fraction by
Meegan et al. (1995), Hartmann et al. (1994b), and Brainerd et al. (1995)
is 20\%.

\section {Dipole and Quadrupole Tests and Distributions}

We test the large-scale isotropy of gamma-ray bursts by
calculating various statistics which measure
deviations from isotropy and comparing
the observed values to the distribution of values
expected if GRBs were isotropic.
If a statistic is found to have a value highly improbable
assuming
GRBs to be isotropic, then the null hypothesis of isotropy is statistically
contradicted.
Several different statistics are used because each is most sensitive to
finding certain patterns of anisotropies.

Large angular scale anisotropies are best searched for with dipole
and quadrupole statistics.   Various dipole and quadrupole statistics
have been introduced to the field of gamma-ray burst studies by
Hartmann \& Epstein (1989), Paczy\'{n}ski (1990) and Briggs (1993) and
are reviewed by Briggs (1993).
Dipole statistics are sensitive to a concentration of GRBs towards one
direction on the sky while quadrupole statistics are sensitive to
concentrations in a plane or towards two opposite poles.   Any plausible
large-scale concentration
should be revealed by its dipole or quadrupole moments before higher
moments become significant.

Dipole and quadrupole statistics based upon a particular
coordinate system are used since they will be most sensitive to
anisotropies in that coordinate system.
Galactic-based tests are used because they are
most sensitive
to Galactic anisotropies, which would result from a Galactic origin for GRBs.
Equatorial-based statistics
are used because
these statistics are most sensitive to the artificial anisotropy induced
by BATSE's nonuniform sky exposure.
We calculate the dipole moment towards the Sun and the quadrupole moment
in the ecliptic plane to test for a heliocentric origin.
A final pair of dipole and quadrupole statistics are constructed in
a coordinate-system independent manner in order to search for anisotropies
in unexpected directions
in a model-independent manner (Briggs 1993).   These statistics are
characterized in Table~1.

Each statistic is used as a one-tailed test: the sign of the deviation
that is significant is indicated in the last column  of Table~1.
We use the coordinate-system
based tests as one-tailed tests because the models tested,
a Galactic origin, a solar system origin, and BATSE's sky exposure,
predict the sign of the expected deviation.
We use the coordinate-system independent tests
${\cal W}$ and ${\cal B}$ as one-tailed tests because no random
anisotropic distribution can cause small values of these statistics.

We use two probability distributions on the sphere in Monte Carlo simulations
designed to test the methods.
A standardized form of the Fisher distribution
has a dipole moment
towards the Galactic center---the probability density  per solid angle is
proportional to $\exp  (\kappa \cos \theta)$, where $\theta$ is the angle
between a point and the Galactic center (Fisher et al. 1987).   Increasing
values of the concentration parameter $\kappa$ result in a greater
concentration towards the Galactic center (Figure~1a).
A standardized form of the Watson distribution has a Galactic-based
quadrupole moment---the probability density per solid angle is
proportional to $\exp  (\kappa \sin^2 b)$, where $b$ is Galactic latitude
(Fisher et al. 1987).
Values of the concentration parameter $\kappa < 0$ result in a concentration
about the Galactic plane, with more negative values of $\kappa$ resulting
in greater concentration (Figure~1b--d).
Values of $\kappa > 0$
result in a concentration at the Galactic poles.
Fisher et al. (1987) and Best \& Fisher (1986) give algorithms for
simulating locations from these distributions
\footnote{Both references have typographic errors in the algorithm for the
Watson distribution.  The algorithm of Fisher et al. (1987) should have
``$\Theta = \arccos S$'' instead of ``$\Theta = S$''.
Also note that this algorithm generates locations only for one
hemisphere---locations must be
moved to the other hemisphere with probability one-half.}.

Figure~2 characterizes these distributions, showing, as a function of
$\kappa$, the moments of the distributions.
Also shown for the Fisher distribution is the angle $\theta_{\frac{1}{2}}$
such that the region within $\theta_{\frac{1}{2}}$ of the Galactic center
contains one-half of the probability.     For the Watson distribution,
Figure~2b shows the angle $b_{\frac{1}{2}}$ such that the region within
$b_{\frac{1}{2}}$ of the Galactic plane contains one-half of the probability.

\section {Analysis}

  \subsection{Location Determination}

BATSE consists of eight detector modules located at the corners of the
Compton Gamma-Ray Observatory.
Each module contains a Large Area Detector (LAD) and a
Spectroscopy Detector (SD).  The results presented here are based upon the
LADs which are 50.8 cm diameter by 1.27 cm thick NaI(Tl) scintillators.
The instrument triggers and records extensive information whenever there
is a 5.5$\sigma$ increase about background in two or more detectors.
Further descriptions of the instrument and the various datatypes
transmitted are given by Horack (1991) and by
Fishman et al. (1994).

BATSE determines GRB locations by comparing the rates in the eight LADs
(Brock et al. 1992a; Fishman et al. 1994).
The locations are determined by the program LOCBURST
by fitting the signal in four detectors
and minimizing $\chi^2$.   The model includes the direct response of the
detectors and scattering in the spacecraft and from the earth.

The locations are not precise because of both statistical
and systematic limitations.
The statistical limitations are due to Poisson fluctuations in the
counts observed with the detectors---these errors $\sigma_{\rm stat}$
are determined by LOCBURST from the derivatives of $\chi^2$, which
are obtained from the observed counts and the detector model.
These errors are well-understood and are Gaussian in the count-regime of BATSE.

Normally there are several datatypes to choose from in order to obtain
the best-possible location, but due to the failure of the CGRO tape recorders,
for 121 GRBs of our total sample of 1005 GRBs
only the MAXBC datatype is available, which consists of the
count rates in each detector for the peak one second of emission.
Because the background is determined differently for MAXBC data than for
other datatypes, until very recently LOCBURST was unable to estimate
statistical
errors for MAXBC-determined locations.

The statistical errors range from a fraction of a degree for the brightest
GRBs to typically $13^\circ$ for GRBs at the trigger threshold, as
summarized in Table~2.
Because the $\sigma_{\rm stat}$ values were unavailable for MAXBC-located
GRBs, these GRBs are not included in Table~2.
For bursts in which the MAXBC datatype must be used to determine the location,
somewhat larger statistical errors are expected: 1) for bursts longer than 1~s,
because of the neglect of source counts outside the peak 1~s interval,
and 2) for bursts shorter than 1~s, because of the reduced SNR caused by
including background-only times in the 1~s long source interval.
After most of the analysis of this paper was accomplished,
LOCBURST was improved to correctly
estimate the statistical errors of MAXBC-determined locations.   Based
upon the currently available sample of 24 reprocessed MAXBC locations,
the statistical error distribution of MAXBC-determined locations is only
moderately worse than that of
locations determined with other datatypes: 14 of the
24 GRBs have $\sigma_{\rm stat} < 3.7^\circ$ and only 3 have
$\sigma_{\rm stat} > 13.8^\circ$, with the largest value being $24^\circ$.

The accuracy of BATSE's GRB locations have been verified in several ways.
For bright GRBs accurately determined locations are available from the
Interplanetary Network (IPN) (Hurley 1993).
Comparison of these locations with the locations determined from the LAD rates
show that the systematic error of the BATSE locations
is currently typically $4^\circ$ (Fishman et al. 1994, Hurley et al. 1994).
These systematic errors are due to calibration uncertainties, imperfect
background subtraction and other
approximations in the data analysis.
We estimate the $1\sigma$ location
error for a GRB as the root-mean-square sum of its estimated statistical error
and a $4^\circ$ systematic error.
Since the $4^\circ$ error is a systematic error, its distribution need not be
Gaussian, and we find from
comparison of LAD-determined locations
with IPN-determined locations that the total error is
approximately Gaussian, with somewhat larger `tails' (Fishman et al. 1994).
The excess tails of the distribution are due to the systematic error component
and are unimportant for bursts for which the statistical error is
dominant.

In addition to their somewhat larger statistical errors $\sigma_{\rm stat}$,
MAXBC-located GRBs should have somewhat larger systematic errors
$\sigma_{\rm sys}$ because the background is determined on-board the
spacecraft by a simpler algorithm than that used by LOCBURST on the ground.
Koshut et al. (1994) analyzed the MAXBC-determined locations of 33 events with
locations determined by the IPN or known from their solar origin and found
that the systematic error was $\lesssim$ $7^\circ$.

The locations have also been verified by examining events from sources
of known location: solar flares, Cygnus X-1 fluctuations and
triggers from SGRs (Fishman et al. 1994; Meegan et al. 1993b;
Kouveliotou et al. 1993a; Kouveliotou et al. 1994a)
These verifications are important since they are not limited to the
bright events required for IPN-locations: solar flares can be of any
intensity and Cygnus X-1 fluctuations and SGRs events are
near threshold.
Fluctuations from Cygnus X-1 are more difficult to locate than GRBs of
comparable intensity because they are always short and are superimposed
on a background of sub-trigger fluctuations.  The distribution
of separations between the location of Cygnus X-1 and the LOCBURST-determined
locations of the fluctuations has a standard deviation of $13^\circ$
(Meegan et al. 1993b),
the typical value of $\sigma_{\rm stat}$ for events near threshold.
For these weak triggers,
the systematic error $\sigma_{\rm sys}= 4^\circ$ is unimportant.

The program LOCBURST has undergone steady improvement: the version used
to determine the post-2B locations used herein is slightly better than
the version used for the post-1B locations of the 2B catalog, which
is distinctly better than the version used for 1B locations.
After the calculation of the locations used herein, major improvements
have been made to LOCBURST which reduce the systematic error
to $\lesssim 2^\circ$.
As we demonstrate below, dipole and quadrupole moments are relatively
insensitive to locations errors so that all of the locations are
useful for this analysis.    The location improvements are important
for tests of clustering or repetition, since the statistical signal
rapidly strengthens with improved locations
(Brainerd et al. 1995; Hartmann et al. 1994b).
At the time of this writing 95 of the locations used herein, all post-2B,
have been redetermined with the newest version of LOCBURST.
Comparison of these revised locations with the locations used herein show that
50\% of the locations have changed by less than
$4.8^\circ$ and  only 10\% have changed by more than $14.0^\circ$.
The largest shift is $28^\circ$.
In terms of the old total error estimates $\sigma_{\rm tot} =$
$ \sqrt{4^2 + \sigma_{\rm stat}}$, 50\% have moved by less
than $0.9\sigma_{\rm tot}$ and 90\% have moved by less than
$2.7\sigma_{\rm tot}$.   The worst shift in terms of $\sigma_{\rm tot}$
is an $18^\circ$ change that is $4.6\sigma_{\rm tot}$.

The anisotropic response of BATSE's detectors,
which enables determination of GRB locations, implies
an intensity threshold for detecting a
GRB which varies across the sky (Brock 1992b).
This intensity-dependent, non-uniform
sky response is best described in CGRO coordinates.   Because of the
many orientations of CGRO for the various observation periods, this
effect averages away and need not be further considered.

Because CGRO is in a low-Earth orbit, about one-third
of the sky is blocked by the
Earth.   This causes the equatorial regions to be observed with about 20\% less
exposure than the polar regions.    Additionally, because
burst triggering is disabled
during passages through the South Atlantic Anomaly (SAA)
and several other regions of enhanced charged particle background,
the South Pole receives about 15\% less exposure than the North Pole.
Both of these effects are intensity independent and are best described in
equatorial coordinates: right ascension, $\alpha$, and declination, $\delta$.
Over the spacecraft orbital precession period of about 50 days, the right
ascension dependence of the exposure averages away, leaving the sky
exposure a function only of declination: $T(\delta)$.   The sky exposure
$T(\delta)$ has been measured (Fishman et al. 1994),
allowing corrections for this effect to be made.

  \subsection{Analysis Methods}

The values of the statistics are calculated from the locations; consequently,
if many samples are ``drawn'' from a population,
the statistics will have different values for each of the samples.
The fluctuations in a finite sample
limit the ability of even a perfect instrument to detect a small anisotropy.
To demonstrate this effect, Monte Carlo simulations of 250 isotropic
locations were made.   Figure~3 shows histograms of the values of
$\langle \cos \theta\rangle$ and $\langle \sin^2 \delta -
\frac{1}{3}\rangle$ obtained from the simulations
overlaid with the theoretically expected asymptotic
distributions (solid curves).
As a characterization of the expected fluctuations, Table~1 lists the
expected standard deviations $\sigma$ of the six statistics, derived from
the asymptotic distributions.
The asymptotic distributions may not be sufficiently accurate
for small numbers of gamma-ray bursts $N_B$ and
extreme values of the statistics,
which is especially likely for
$\langle \sin^2 b - \frac{1}{3}\rangle$ (Briggs 1993).
When necessary
we use the actual distributions of the statistics as determined
by Monte Carlo simulations of isotropy.

An artificial anisotropy should exist in the observed sample of GRB locations
because BATSE's sensitivity is not uniform across the sky.
The resulting bias expected in the statistics has been
been found by Monte Carlo
simulations---isotropic locations are created using a random number
generator and included in the simulated sample
with a probability obtained from the sky exposure map $T(\delta)$.
Locations are generated until $N_B$ have been accepted
and then the
values of the statistics are calculated.
This effect is illustrated in Figure~3 by the second histograms with their
overlying dashed curves.   The histograms are still Gaussian, but the
artificial anisotropy of the sky exposure has caused the expected mean
of the observation to shift from zero.   In addition, there is a very small
change in the widths of the distributions.
The shift in the mean is large for $\langle \sin^2 \delta - \frac{1}{3}\rangle$
because this equatorial-based statistic is sensitive to the time-averaged
shadowing of the celestial equator by the Earth.
The shift in the mean of $\langle \cos \theta \rangle$ is small because
the dipole induced by the equatorial North-South asymmetry is a smaller effect,
relative to the standard deviation $\sigma$ of the fluctuations,
than the quadrupole moment induced by the Earth blockage and because
only part of the equatorial dipole is projected onto the direction towards
the Galactic center.
The fraction of the artificial dipole moment projected equals the cosine of
the angle between the equatorial plane
and the Galactic plane, which is $\cos (60^\circ)$ = \slantfrac{1}{2}.

Table~3 lists
the means and standard deviations of the eight
statistics expected for isotropy modified by BATSE's actual sky exposure.
These sky-exposure-corrected mean values of the statistics have been
determined using BATSE's sky exposure accumulated
for the time period of the first BATSE catalog (Fishman et al. 1994),
the 320 days from 1990 April 21 to 1992 March 5.
Work is in progress to measure the sky exposure after this period, a task
made difficult by the failure of the CGRO tape recorders.
The expected values have also been calculated
for the sky exposure map for the first 165.5 days (Brock et al. 1992b)---the
differences are insignificant.    We expect the changes in BATSE's sky
exposure with time to remain small, so the extrapolation of the current
sky exposure map to $N_B=2000$ gamma-ray bursts should be a good approximation.

These two effects, the finite sample fluctuations and the shifts in the
expected means due to BATSE's sky exposure, are illustrated in Figure~4.
The dashed lines show the results expected ignoring BATSE's sky exposure,
while the solid lines indicate the results expected based upon BATSE's
sky exposure.    The bold lines show the expected means of the statistics
while the normal-weight lines show the $\pm1  \sigma$ envelopes of the
distributions of the statistics.   The widths of the distributions (indicated
by the outer lines) are due to the finite sample fluctuations, while the
shifts of the means (indicated by the separation between the
dashed, bold lines and the solid, bold lines) are due to the
artificial anisotropy induced by BATSE's sky exposure.

The lines on Figure~4 allow visual evaluation of
the effects of BATSE's nonuniform sky exposure.
We consider the effect of the nonuniform sky exposure to be major when
it causes the corrected mean of a statistic
to differ from the ideal mean by $\gtrsim 1  \sigma$,
i.e., when the bold, solid line is outside of the outer dashed lines.
For the Galactic-based statistics, $\langle \cos \theta \rangle$ and
$\langle \sin^{2} b - \frac{1}{3} \rangle$, this only happens for
$N_B \gtrsim 2000$.
These statistics are relatively insensitive to the nonuniform sky exposure
because the Earth's equator is highly inclined to the Galactic equator.
The coordinate-system independent quadrupole statistic ${\cal B}$ is not
importantly affected until $N_B \gtrsim 400$,
while the coordinate-system
independent dipole statistic ${\cal W}$ is not importantly
affected by the nonuniform sky exposure until $N_B \gtrsim 1200$.
Because the nonuniform sky exposure originates in the equatorial coordinate
system, the equatorial-based statistics are quite sensitive to it:
the effects are important for $\langle \sin \delta \rangle$ for
$N_B \gtrsim 500$ locations and for
$\langle \sin^{2} \delta - \frac{1}{3} \rangle$ for $N_B \gtrsim 130$
locations.
The dipole statistic $\langle \sin \delta \rangle$ detects the $\simeq$
15\% increased exposure of the North Pole relative to the South Pole, while
the quadrupole statistic $\langle \sin^{2} \delta - \frac{1}{3} \rangle$
detects
the $\simeq$ 20\% reduced exposure of the equator relative to the poles.

As described above, the locations of the GRBs are uncertain, which
propagates into uncertainties in the values of the statistics.
Because the location errors are small compared to the scale of large-scale
anisotropies, the corresponding errors in the values of the statistics
are negligible.
Monte Carlo simulations were conducted to
demonstrate this.
The histograms on the left of each panel of Figure~5
are based upon $5\times 10^5$ simulations of 1000
isotropic locations.
The curves overlaying these histograms are the asymptotic
distributions of the statistics.
The vertical lines on Figure~5 show the critical values of the statistics
corresponding to confidence levels against the null hypothesis of isotropy of
99\%, 99.9\%, and 99.99\%.
These critical values are defined by the requirement that the chance
probabilities
under the assumption of isotropy
of more extreme values of the statistics are
1\%, 0.1\%, and 0.01\%.

Simulations were made of two distributions with
moderately detectable anisotropies.
The results of the simulations made assuming perfect determinations of the
locations are shown in the bold, right-most histograms on Figure~5.
The fractions of these histograms to the right of the vertical lines
are the fractions of the simulations in which the anisotropies are detected at
the confidence levels of the lines.
The anisotropic simulations were repeated, this time
moving each simulated location an angle $\Sigma$ in a random direction,
obtaining the middle histograms of Figure~5.

The angles $\Sigma$ and the fractions of the simulated bursts moved by
each particular value of $\Sigma$
are listed in the last two columns of Table~2.
The simulated smearing exaggerates with a very substantial
safety margin the actual location uncertainties:
1) the values $\Sigma$ are based upon the worst $\sigma_{\rm tot}$
in each row, 2) the values $\Sigma$ are at least twice this worst location
error,
3) the simulated smearing locations are each moved by an angle $\Sigma$,
{\it not} by an angle selected from
a Gaussian distribution of this width, and 4) to allow for outliers
additional simulated locations are placed into the category with the
largest smearing angle.

Figures~5a~\&~b show simulations of the Fisher distribution with $\kappa=0.2$,
a slightly more concentrated distribution than the example shown in Figure~1a.
This distribution places one-half of the probability
within $84.3^\circ$ of the Galactic center and has an expected dipole
moment of $\langle \cos \theta\rangle$ = 0.06649.
The middle histograms, which show the values
of the statistics
when the locations are smeared, are only slightly shifted from the bold
histograms on the right, which show the results for perfect locations.
Because the
scale of the anisotropy is of order $90^\circ$,  the smearing by tens of
degrees is a small effect.
To test the effect
of the location errors for a quadrupole distribution, simulations of the Watson
distribution for $\kappa=-10$ (Figure~1d) were made---this
distribution has
one-half of the probability within $| b | \leq 8.7^\circ$
and an expected quadrupole moment of
$\langle \sin^{2} b - \frac{1}{3} \rangle$ = $-0.2833$.
This
distribution is extremely detectable, so its detectability was reduced by
diluting it with isotropic locations, creating anisotropic locations with
probability 10\%.   In this example (Figures~5c~\&~5d), the middle
histograms based upon the smeared
locations are quite noticeably shifted from the bold histograms based upon
the perfect locations, but the smearing has not caused the middle
histograms
to be similar to the histograms for isotropic
locations.

The sensitivity or power of a statistic is the probability of
its finding an actual anisotropy (Eadie et al. 1971; Martin 1971).
In our case, the powers of the statistics for finding the example anisotropies
at the confidence levels
of 99\%, 99.9\% and 99.99\%
is given by the fraction of simulations with statistics more extreme
than the corresponding critical values, i.e., to the right of the vertical
lines in Figure~5.
The powers of the statistics are listed in Table~4.
The decrease in the fraction of the anisotropic simulations, smeared versus
unsmeared, that are past the critical values represents the loss in
sensitivity or power to the anisotropy caused by the smearing.
Table~4A shows that the loss of power for the Fisher $\kappa=0.2$
model is minor.
Similar results were obtained by Horack et al. (1993) via simulations
of a halo distribution.
The model with 90\% isotropic locations and  10\% locations from a narrow disk
model was intended
to be an extreme case---half of the disk probability is contained within
$| b | \leq 8.7^\circ$ versus the exaggerated smearing angles $\Sigma$
listed in Table~2.
Table~4B shows that
while the smearing substantially reduces the sensitivity to this
model anisotropy, the ability to detect the narrow disk component is not
eliminated.

The above simulations were each of 1000 locations.
We now discuss how our ability to find anisotropies with BATSE depends on
the number of gamma-ray bursts $N_B$ observed.
The finite-sample fluctuations of the coordinate-system based statistics,
such as $\langle \cos \theta\rangle$ and
$\langle \sin^{2} b - \frac{1}{3} \rangle$, asymptotically have a
$1/\sqrt{N_B}$ dependence
(Tables 1 \& 3) because of the Central Limit Theorem (Paczy\'{n}ski 1990;
Briggs 1993).
Monte Carlo simulations show that the asymptotic distributions
are sufficiently accurate for the values of $N_B$ under consideration
(Briggs 1993).
The coordinate-system independent Rayleigh-Watson
${\cal W}$ and Bingham ${\cal B}$ statistics are proportional to the
product of $N_B$ and the moment per GRB squared (Bingham 1974;
Watson 1983; Briggs 1993) and their asymptotic
distributions are also sufficiently accurate (Briggs 1993).
Thus for both types of statistics the
finite sample fluctuations cause the magnitude of detectable
anisotropic moments to scale as $1/\sqrt{N_B}$.
The effects of the location uncertainties scale in the same manner.
It doesn't matter where a variance comes from, only that it is finite,
in which case
the Law of Large Numbers or the Central Limit Theorem tells us that the
mean of the measurements will be determined with uncertainty scaling as
$1/\sqrt{N_B}$ for large $N_B$ (Eadie et al. 1971; Martin 1971;
Lyons 1986).
This is the well-known result that the knowledge of the mean of a series
of numbers with errors improves as $1/\sqrt{N}$.

As a demonstration that BATSE's ability to find anisotropies improves as
$1/\sqrt{N_B}$, $5\times 10^4$ simulations
of $10^4$ locations were made (Tables 4C and 4D).
The simulations in Tables 4C and 4D have 10 times the number of locations
as those of Tables 4A and 4B, but the moments have been reduced by a
factor of $\sqrt{10}$.
Allowing for the fluctuations expected in $5\times 10^4$ simulations, the
anisotropies are detected in the same fraction of simulations in
Tables 4C and 4D as in Tables 4A and 4B, confirming the $1/\sqrt{N_B}$
improvement in sensitivity.

Smearing by angles $\Sigma$ as listed in Table~2 reduces the dipole
moment $\langle \cos \theta \rangle$ of dipole distributions by
a factor of 0.86 and the quadrupole moment
$\langle \sin^2 b - \frac{1}{3} \rangle$ of quadrupole distributions
by a factor of 0.68.   We found this to be true not only of the
distributions listed in Table~4, but also for all the other distributions
we simulated.    These reduction factors greatly overestimate the effects
of BATSE's location uncertainties since the angles $\Sigma$ greatly
exaggerate the location uncertainties.

Another possible error in the values of the statistics is due
to the possibility of trigger misclassifications.
We estimate both the number of GRBs misclassified as non-GRBs and the number
of non-GRBs misclassified as GRBs to be about 1\%
(Mallozzi et al. 1993; Fishman et al. 1994).
The estimated misclassifications are roughly equally split between
Cygnus X-1 fluctuations and solar flares,
so the misclassifications are not all at the same location.
We also estimate that at most one Soft Gamma-Ray Repeater (SGR)
has been misidentified as a GRB and
vice versa (Kouveliotou 1994b).
Even if all 9 SGR triggers are added
to the set of 1005 gamma-ray GRBs which we analyze, the set remains
consistent with isotropy.
In an extreme example, if a false population of 2\% of the GRBs is placed
at the Galactic Center, a dipole moment of
$\langle \cos \theta \rangle = 0.02$ is produced,
which would be detectable only at the 1.1$\sigma$ level in a sample of
1000 locations.

In summary,
the dominant limitation for BATSE in discovering a large-scale anisotropy
is the fluctuations resulting from the finite number of GRBs observed, a
limitation that no instrument can avoid.   The effects of the location
uncertainties are negligible and will remain so since
the effects of both the finite-sample
fluctuations and the location uncertainties scale as
$1/\sqrt{N_B}$.
The estimated rate of trigger misclassifications is too low to significantly
effect the results.
The bias induced by BATSE's nonuniform
sky exposure can be measured and corrected for
when comparing model predictions with the observations.

\section {Results}

The first 1005 gamma-ray bursts observed by BATSE are shown in Figure~6.
These GRBs occurred between the enabling of burst triggering on
1991 April 21 and the end of CGRO Viewing Period \#222 on 1994 May 31.
The post-2B catalog locations
are preliminary, but as discussed in \S3.1, the revised locations will
be close to the current locations and are
adequate for the study of the large-angular scale properties of
gamma-ray bursts.  The 585 locations of the second BATSE catalog (2B) are
publicly available (Meegan et al. 1994b).   The first 262 triggers of the
2B catalog are referred to as the revised 1B catalog---these differ from
the original 1B catalog (Fishman et al. 1994) by the revised classifications of
several triggers.

The values of the six statistics for the first 40, 100, 262 (= 1B revised),
585 (= 2B) and 1005 gamma-ray bursts
observed by BATSE are shown on Figure~4 with dots (note that these sets are
not independent since each is a subset of the latter sets).
A greater than $1  \sigma$ deviation of a statistic from the value expected
for isotropy is indicated by the corresponding dot lying outside of
the outer solid lines, which represent the $\pm1  \sigma$
envelopes of the distributions of the statistics, including the effects of
BATSE's sky exposure.   This occurs several times, but only by a small
margin.    Thus the depicted statistics are all consistent with isotropy.

In order to test BATSE's sky exposure map,
Table~5 presents the Earth-based statistics, $\langle \sin \delta\rangle$ and
$\langle \sin^2 \delta - \frac{1}{3} \rangle$, for the revised 1B catalog,
the 2B catalog and the first 1005 GRBs.   The largest deviation from
the predictions based upon the sky exposure map is $1.3\sigma$; the
largest deviation
for the first 1005 GRBs is only $0.7\sigma$.
In contrast, the value of $\langle \sin^2 \delta - \frac{1}{3}\rangle$ for
the 1005 GRBs
is $2.0\sigma$ away from the mean that
would be expected for a uniform sky exposure.
This statistic is the most sensitive to BATSE's nonuniform sky exposure and
appears to have detected that effect at modest significance.
These results are good indications
that we are accurately predicting the effects of BATSE's nonuniform sky
exposure.

Table~6 lists the subsets into which the set of 1005 GRBs have
been split in order
to test for possibly anisotropic subsets.
Many of these sets have common members and are thus not independent.
For each of these sets the Galactic and coordinate-system independent
statistics have been calculated and are listed in Table~7.

Because Table~7 lists the values of four statistics for 48 subsets of the
1005 GRBs, it is likely that at least one value
of a statistic will
have a large deviation from the predictions of the null hypothesis of isotropy.
The four statistics are not completely independent, just as the
various subsets are not all independent.
Ultimately the degree of
belief corresponding to a confidence level is a scientific judgement.
In our judgement,
for the entire dataset an improbability of
$\lesssim 0.01$
would be interesting and
$\lesssim 0.001$ would be convincing.
Because of the many subsets examined, we would decrease these thresholds
by at least a factor of 10 when considering a candidate anisotropy
present only in a subset.

Of the 48 datasets analyzed in Table~7, only one deviates from isotropy at
more then the 99\% confidence level,
namely dataset \#46, the Medium GRBs of the
revised 1B catalog.
These 51 GRBs are ``Medium'' according to the selection criteria of
Quashnock \& Lamb (1993a): $\log_{10} V < -0.8$ and $465 < B < 1169$, where
B is a measure of GRB brightness on the 1024 ms timescale and V
compares the peak counts rates on the 64 ms and 1024 ms
timescales (Lamb, Graziani \& Smith 1993).
This dataset deviates from isotropy by $+2.4\sigma$ for
$\langle \cos \theta\rangle$ and by $-2.6\sigma$ for
$\langle \sin^2 b - \frac{1}{3}\rangle$.
These deviations are in the direction of a concentration towards the
Galactic center and the Galactic plane.
The dataset of Quashnock \& Lamb (1993a) was selected from the 1B catalog
and had four additional GRBs;
the $C_{\rm max}/C_{\rm min}$ values of these four GRBs
on untriggered timescales have been
withdrawn by the BATSE team as being erroneous and these four GRBs are
therefore
not included in the Medium bursts of the revised 1B catalog.
The 55 GRB dataset of Quashnock \& Lamb (1993a)
deviates from the predictions of isotropy and
BATSE's sky exposure map
by $+3.2\sigma$ for
$\langle \cos \theta\rangle$ and by
$-2.9\sigma$ for $\langle \sin^2 b - \frac{1}{3}\rangle$.
The 51 Medium GRBs
of the revised 1B catalog are shown with filled circles on Figure~7 while the
additional four from the original 1B catalog are shown with open circles.

The dipole and quadrupole moments of the Quashnock \&
Lamb 55 GRB dataset each have significances of roughly $10^{-3}$.
Quashnock \& Lamb
point out that these two statistics are partially independent and find from
Monte Carlo simulations that the probability of seeing both deviations
or larger is $1.9\times 10^{-6}$.
Because they optimized the brightness B selection criterion to obtain this
result, they estimate
the probability as $1.1\times 10^{-4}$.
They interpreted their dataset (filled and open circles on Figure~7) as
evidence
for spiral arms and concluded that all gamma-ray bursts are Galactic in
origin (Quashnock \& Lamb 1993a).

The estimation of the correction factor to apply to significances obtained from
retrospective analysis of data is difficult.
How many ways were the data subdivided?    How many statistical approaches
were tried?    In this case, we should have some additional factor
due to the V selection criterion and to the idea of using a joint
probability from $\langle \cos \theta\rangle$ and
$\langle \sin^2 b - \frac{1}{3}\rangle$.
While retrospective analysis has potential pitfalls, it is
necessary because of our limited understanding of gamma-ray bursts.

The best solution is to collect more data in order to make an independent
test of a hypothesis formed by retrospective analysis.
Using exactly the same selection criteria as Quashnock \& Lamb
(Graziani 1993), we find 76 additional Medium GRBs in the
first 1005 GRBs---these
form our dataset \#47 and are shown with stars on Figure~7.
The rate of discovering Medium GRBs has decreased due
to data gaps caused by the failure of the CGRO tape recorders which prevent
our determining the B and V values for many GRBs.
Even a small data gap, which does not hinder GRB localization,
prevents determination of the peak count rate because
the peak rate may have occurred during the data gap.
The
data gaps should not have any correlation with intrinsic properties of
gamma-ray bursts and thus should not bias the selection.
The new Medium GRBs are consistent with isotropy and in fact the
$-1.7\sigma$ deviation of $\langle \cos \theta\rangle$ indicates
a {\it deficiency} of GRBs towards the Galactic center.
The evidence for a Galactic origin of GRBs is not confirmed by
a larger, independent sample.
The entire dataset of 51 + 76 = 127 Medium GRBs, dataset \#48,
is also consistent with isotropy:
$\langle \cos \theta\rangle$ deviates by $+0.2\sigma$ and
$\langle \sin^2 b - \frac{1}{3}\rangle$ by $-1.8\sigma$.

What can we conclude from the contradictory results of datasets \#46 and 47?
If one takes dataset \#46 to be evidence of a true anisotropy and thus
the values of
$\langle \cos \theta\rangle$ and
$\langle \sin^2 b - \frac{1}{3}\rangle$ observed for this set
to be measurements of true anisotropic moments,
then one has the difficulty of explaining the $3.0\sigma$ and and $1.8\sigma$
deviations of the moments observed for dataset \#47 from these values.
Conversely, if one assumes the GRBs to be isotropic, then one has the
difficulty of explaining the
$+2.4\sigma$ and the $-2.6\sigma$ of deviations of
$\langle \cos \theta\rangle$ and $\langle \sin^2 b - \frac{1}{3}\rangle$
for dataset \#46 from the values expected for isotropy.
Because dataset \#46 was identified by retrospective
analysis and its
result is not confirmed by a larger, independent sample, we judge that
the evidence against the null hypothesis of isotropy is weak.

We use the statistics of Horack et al. (1994) to search for a heliocentric
pattern in the first 1005 GRBs.    The dipole moment towards the Sun,
$\langle \cos \phi \rangle$, is observed to be $0.006$, which deviates by
$+0.3\sigma$ from the value of $0.000$ predicted by isotropy and
BATSE's sky exposure.   Because the direction to the Sun changes with time, the
sky exposure bias is calculated by averaging
the time-dependent bias specified in Table~3 over 3.1 years.
The quadrupole moment in the ecliptic plane,
$\langle \sin^2 \beta - \frac{1}{3} \rangle$, is observed to be
0.022, a $+0.2\sigma$ deviation from the value of $0.020$ predicted
by isotropy and BATSE's sky exposure.
No evidence for a heliocentric origin is found.

We have searched for an excess or deficiency of GRBs near the
Magellanic Clouds, the Virgo cluster, M31, the Galactic Center, and
$\alpha$ Centauri.   In all 48 datasets there is no instance
in which the number of GRBs within $20^\circ$ of these
objects is less probable than 1\%.

In summary, we have examined 48 subsets of the first 1005 gamma-ray bursts
observed by BATSE using four dipole and quadrupole tests.
The entire set of 1005 GRBs has been tested with four additional
dipole and quadrupole statistics.
Only one set, \#46, appears anisotropic, however, a
larger, independent set selected with the same criteria, \#47, fails to confirm
the anisotropy so we judge the data to be consistent with isotropy.
The largest anisotropy in any set other than \#46 is the $2.1\sigma$
deviation of $\langle \cos \theta \rangle$ for  dataset \#27.    Because
of the large number of partially independent sets and statistics examined,
a deviation of this magnitude is expected.
The gamma-ray bursts locations are consistent with large-scale isotropy.

\section{Model Comparisons and Limits}

While the statistics show that gamma-ray burst locations
are consistent with isotropy, they do not prove that gamma-ray bursts
are isotropic---one could imagine that a small anisotropy exists that cannot
be detected in a sample of 1005 locations.
This raises the question of how large an anisotropy could exist and
not have been detected.
We address this question in two ways:
First, we discuss below and summarize in Table~8 the comparisons of
the moments observed by BATSE with the moments of published
Galactic models.
Secondly, we use Monte Carlo simulations to
determine for a variety of models which anisotropies could have been detected.
These upper-limits on anisotropies are discussed below and summarized in
Figure~8.

For both the model comparisons (Table~8) and model parameter limits
(Figure~8) we have assumed that the number of sources
is the same as the number of observed bursts.   If some of the sources
repeat so that there are fewer sources than observed bursts, the finite
sample fluctuations will be larger than we have calculated, which will
reduce our ability to detect anisotropies (Quashnock 1995).
We have placed a 20\% upper-limit on the repeater fraction at the
99\% confidence level (Meegan et al. 1995),
which implies that there are at least 9 sources for each 10 observed
bursts.
In this case the finite sample fluctuations would be 10\% larger than
we have calculated.

\subsection{Model Comparison Methods}

Below and in Table 8 we compare the moments of published Galactic models
with the observations.
Because the goal of these comparisons is to confront the models with
BATSE's observations, the comparisons are made between the observed
moments corrected for the sky exposure bias and the model moments.
These published models were developed based upon earlier and smaller
datasets---some could have improved agreement with the current observations
by reoptimizing their parameters.
In other cases, full reoptimization might increase the deviations in the
moments because of the tighter constraints of the brightness distribution
data.
Some of the models are merely examples of good fits to older data, rather
than best-fits to that data.
These models might be improved to have reduced moments, possibly at the
expense of less ``reasonable'' parameters.
The uncertainties in the theoretical models are difficult to quantify.
In any case, reoptimizing the models is beyond the scope of this paper, so we
compare the observations with the published moments without any corrections.
We convert the differences between the observed and published model
moments to deviations in units of $\sigma$, using for $\sigma$
the finite sample fluctuations of the observations,
ignoring the possibility of uncertainties in the models.

The values $\sigma$ are calculated as the sample fluctuations expected
assuming isotropy (Table~3)---a more accurate procedure would
be to calculate the values of $\sigma$ from the anisotropic models (Li 1995).
For models with moments so small that $\sim 1000$ bursts are needed to
detect the moment, the difference between the model and isotropy $\sigma$
values are typically $\lesssim 1 \%$.
For the model of Li et al. (1994), which predicts a larger moment in
a subset of the data, the difference somewhat exceeds 10\% (Li 1995).
Since the differences between the isotropy and model $\sigma$ values are small
and the model $\sigma$ values generally have not been published, we calculate
the $\sigma$ values assuming isotropy.

In \S3.2 we found that smearing the locations by the angles $\Sigma$ listed
in Table~2 reduced the dipole moment $\langle \cos \theta \rangle$
by the factor 0.86 and the quadrupole moment
$\langle \sin^2 b - \frac{1}{3} \rangle$ by 0.68.   These factors
considerably exaggerate the effects of smearing since the angles
$\Sigma$ are chosen to be at least twice $\sigma_{\rm tot}$
for the least well-located GRB
in each of four categories (Table~2).    Simulations using more accurate
smearing angles show the smearing factors to be only a few percent below
one, and we have therefore neglected this effect in our model
comparisons (Table~8).

\subsection{Model-Limit Determination Methods}

Monte Carlo simulations are used to determine for various models,
as a function of the number of bursts $N_B$,
the parameter values that should create detectable anisotropies.
The goal of the simulations is to determine what anisotropies should
be detectable and therefore the upper-limits are not based upon the
moments observed by BATSE.

In order to calculate the ability of a statistic to detect an anisotropy,
the ``power'' of a statistic,
it is necessary to
assume a particular anisotropic model (Eadie et al. 1971; Martin 1971).
For example, the statistic $\langle \cos \theta \rangle$ is highly
sensitive to a dipole moment towards the Galactic center but completely
insensitive to a dipole moment towards the North Galactic Pole.
For each model
we determine the detectability of anisotropies for three confidence levels,
99\%, 99.9\% and 99.99\%, which are equivalent to
$2.3\sigma$, $3.1\sigma$, and $3.7\sigma$ for Gaussian-distributed
statistics.
The critical values of the statistics are determined from the
distribution of the statistics under isotropy so that the chance probability
of mistakenly deciding that an isotropic distribution is anisotropic is
1\%, 0.1\%, and 0.01\% (see the vertical lines on Figure~5).
Choosing a high confidence level reduces the chance of
falsely discovering an anisotropy, a mistake known as a Type I error,
at the expense of reducing the ability to discover a true anisotropy
(Eadie et al. 1971; Martin 1971).
We show the three confidence levels to allow readers to choose based upon
their judgement and considerations such as whether the entire dataset
or a subset is being tested.

Just as an isotropic distribution may by chance look rather anisotropic,
so may an anisotropic distribution appear rather isotropic.
This is seen in Figure~5 in the areas of the anisotropic histograms that
extend near the center of the isotropic histograms.
A Type II error is failing to discover an anisotropy when it actually exists
and the power of a statistic is one minus the probability of making a
Type II error, i.e., the probability of detecting the anisotropy
(Eadie et al. 1971; Martin 1971).
We choose to determine the detectability of anisotropies for two powers:
for BATSE ``typically'' and ``nearly always''
discovering an anisotropy, by which we mean
finding the anisotropy at or past the stated confidence level in 50\%
and 95\%, respectively, of the simulations.

Figure~8 shows as a function of the number of gamma-ray bursts $N_B$ the values
of the anisotropy parameters such that either 50\% or 95\% of the simulations
find the anisotropy at one of our confidence levels.
The detectabilities are determined as a function of $N_B$ for
$N_B$=100 to 2000.   This allows one to determine what anisotropy could
be found in a subset of the current 1005 GRBs or will be detectable
in the future.
Each panel of Figure~8 thus has six curves: three confidence levels at
two different powers.
The ``typical'' detection curves are solid while the ``nearly always''
curves are dashed.   The y-axes are arranged so that more
anisotropic models are at the top of the graph, hence, the top curves are
for the 99.99\% confidence level and the bottom curves are for the 99\%
confidence level.

We note that in all cases detecting an anisotropy ``nearly always'' at the
99\% confidence level (lowest dashed curves) requires a larger anisotropy
than detecting the anisotropy ``typically'' at the 99.99\%
confidence level (highest solid curve).
This is an indication
of how conservative in the sense of minimizing BATSE's ability to
find anisotropies the ``nearly always'' requirement is.

The detectability of each model is determined for the most sensitive
Galactic statistic for that model (left panels) and for the most sensitive
coordinate-system independent statistic (right panels).    This allows
one to determine what Galactic anisotropies and what
anisotropies in less-expected directions, e.g., M31, Virgo cluster, Magellanic
Clouds, BATSE could detect.
With one exception discussed below (\S5.4), the Galactic-based statistic are
more sensitive for discovering our example models, which are constructed
in Galactic coordinates.

The location errors slightly ``isotropize'' the observations.
The Monte Carlo simulations used to determine the detectability of the
model anisotropies (Figure~8)
include a ``smearing'' of the simulated locations by an angle
$\sigma = 14^\circ$.
The value $14^\circ$ is the root-mean-square
sum of the $13^\circ$ statistical and $4^\circ$ systematic error typical
for BATSE's faint GRBs and is therefore conservative in the sense of
underestimating our ability to find anisotropies.

\subsection{Fisher and Watson Models}

We use the Fisher and Watson distributions as ``generic'' dipole and
quadrupole distributions to derive some illustrative limits.
We assume that all locations generated in the simulations of these
distributions are ``observed''.
Figures 8a and 8b
show the detectable values of the concentration parameter $\kappa$ for these
two models.    Values of $\kappa$ may be converted into dipole and quadrupole
moments and into angles containing one-half of the probability using
Figure~2.

The lowest curve of the left box of
Figure~8a shows that for 1000 GRBs BATSE would ``typically'', i.e.,
50\% of the time, detect a Fisher
distribution with $\kappa \geq 0.13$ at the 99\% confidence level or higher;
a simulated example of this distribution
is shown in Figure~1a.
The dipole moment $\langle \cos \theta \rangle$ of this
distribution is 0.043, implying an expected significance for 1000 GRBs
of $2.4 \sigma$.
The Fisher distribution with
$\kappa = 0.13$ has one-half of the probability within
$\theta_{\frac{1}{2}} = 86^\circ$ of the Galactic center
compared to $90^\circ$ for isotropy.

A very conservative upper-limit obtained from the middle dashed curve of
the left box of
Figure~8a is that BATSE would with 1000 locations
``nearly always'', i.e., 95\% of the time,
detect a Fisher distribution about the Galactic center with
$\kappa \geq 0.27$ at the 99.9\% confidence level or better.
This distribution has one-half of the
probability within $\theta_{\frac{1}{2}} = 82^\circ$ of the Galactic center
and an expected
moment $\langle \cos \theta \rangle$ = 0.090, implying a typical detection
for 1000 GRBs
at the very high level of $4.9 \sigma$.

Similar results for finding a quadrupole moment created by the Watson
distribution are given in Figure~8b.   A value of $\kappa = -0.28$
would typically be discovered in 1000 locations by the
$\langle \sin^{2} b - \frac{1}{3} \rangle$ test
at the 99\% confidence level (see Figure~1b).   This
distribution has one-half of the probability within
$b_{\frac{1}{2}} = 28^\circ$ of the Galactic plane versus $30^\circ$
for isotropy.  The fact that a plane concentration this weak is detectable
but not detected shows how isotropic the GRBs are.
The middle dashed curve of the left box of Figure~8b gives very conservative
upper limits, e. g., for 1000 GRBs, BATSE would nearly always find
at the 99.9\% confidence level a Watson distribution with $\kappa=-0.57$.
This distribution has one-half of the probability $26^\circ$ of the
Galactic plane.

The right box of Figure~8a gives
results for detecting a concentration about any point on the sky: a value of
$\kappa \geq 0.18$ would be typically detected in 1000 GRBs
by the Rayleigh-Watson
dipole statistic ${\cal W}$ at the 99\% confidence level.
The corresponding concentration is having one-half of the probability
within $85^\circ$ of any point.
Similarly the right box of Figure~8b gives results for detecting a
concentration about any plane on the sky:
a value of $\kappa \leq -0.4$ would be typically detected in 1000 GRBs
by the Bingham quadrupole statistic ${\cal B}$ at the 99\% confidence level.
The corresponding concentration is having one-half of the probability
within $27^\circ$ of some plane.
These results constrain suggestions that gamma-ray bursts originate from
the Magellanic Clouds (Maoz 1993a; Fabian \& Podsiadlowski 1993)
or from the Milky Way and M31 (Gurevich et al. 1994)
or from any other unexpected direction on the sky.

\subsection{Disk and Spiral Arm Models}

The Watson distribution discussed above serves
as a non-physical model of a disk distribution.
We also consider a somewhat more physical model: the GRBs are assumed to
be standard candles distributed exponentially in distance from the Galactic
plane and the solar system is assumed to lie in the plane.
Note that the vertical profile of the thin stellar disk of the Milky Way is
adequately modeled by an exponential
(Gilmore, King \& van der Kruit 1990).
The appearance of this distribution is parameterized by the ratio
observing depth, which is the maximum distance to which GRBs can be seen,
divided by the exponential scale height.
For 1000 GRBs, even using the very conservative limit of nearly always
detecting
the anisotropy at the 99.9\% confidence level, the middle dashed curve of
Figure~8c left
shows that BATSE cannot be viewing past 0.8 exponential scale heights.
In this model BATSE is viewing a spherical region no larger than 0.8
scale heights and the ratio of the lowest density of GRBs viewed over the
highest density is $\exp (-0.8) = 0.45$, which is reached only at the
North and South Galactic Poles of the spherical region.
Since the faint GRBs are actually deficient by a factor of about six,
this model clearly does not agree with the inhomogeneous radial distribution
observed by BATSE.

As a very simple spiral arm model we consider an infinitely-long cylinder
containing a uniform density of standard candle sources,
with the Sun located on the cylinder axis.    This
distribution is parameterized by the ratio observing depth divided by
arm radius.
Using the very conservative upper limit of nearly always detecting
the anisotropy at the 99.9\% confidence level, the middle dashed curve of
Figure~8d right shows that the observing depth cannot exceed 1.2 arm
radii.
The fractional volume of the sphere in which GRBs could be seen occupied by
the uniform density cylinder is 83\%.
Again, this clearly does not match the major deficiency of
faint GRBs observed by BATSE.

Only in the spiral arm model is
the coordinate-system independent statistic more
sensitive than the Galactic-based statistic.
This is because $\langle \sin^{2} b - \frac{1}{3} \rangle$ is only detecting
the concentration the arm makes in the plane, while the Bingham statistic
${\cal B}$ is also detecting the concentration of GRBs at the two poles
of the cylinder axis.    We could construct a special Galactic-based
statistic to look for this latter effect, but the limits obtained with
${\cal B}$ amply suffice.

In both the disk and spiral arm models one would expect the faintest sources
to show the largest anisotropy, but no anisotropy is detected in these
GRBs, e.g., datasets \#21, 28, 35.

It should be clear that a disk or spiral arm model
requires a strong anisotropy because of the major deficiency of faint
GRBs seen by BATSE.    For example, in a disk model, the many faint GRBs
which are
``missing'' above and below the plane in order to create the observed
inhomogeneity would necessarily produce a concentration in the plane and
a large quadrupole moment.
This was shown to be the case for several disk models by
Paczy\'{n}ski (1990, 1991b) and has been confirmed for a variety of models
by Mao and Paczy\'{n}ski (1992b) and Hakkila et al. (1994a).
The inability of spiral arm models to fit the BATSE data has been shown
by Hakkila et al. (1994a) and by Smith (1994a).
A model with more than one spiral arm visible would be subject to the
constraints against an excess in the plane characteristic of disk models.

\subsection{Geometric Halo Models}

The basic requirements on a halo model are:
1) the length scale of the distribution should be
$\gg$  the solar galactocentric distance
$R_\circ = 8.5 {\rm kpc}$,
so that the distribution will appear sufficiently isotropic,
2) the observing depth (the maximum distance seen by BATSE) should be $\gg$
the length scale of the distribution
so that there will be a sufficient deficiency
of faint GRBs, i.e., inhomogeneity, and 3) the observing depth should
be $\lesssim$ 350 kpc, half the distance to M31, so that the model will not
predict
an excess of GRBs from M31.    Since the focus of this paper is large-scale
isotropy, we will primarily be concerned with the first requirement:
that our offset from the center of the
galaxy be small compared to the size of the observed GRB distribution so
that the dipole moment not be too large.   Generally the constraint
imposed by the quadrupole moment is less limiting.

Any spherically symmetric
halo model may be thought of as the sum of a series of shells.
Hartmann et al. (1994a) have shown that a shell of galactocentric
radius R has a dipole moment
\begin{equation}
\langle \cos \theta \rangle =\frac{2}{3}\frac{R_\circ}{R}.
\end{equation}
Thus a shell of radius $R=110$ kpc has a dipole moment $2\sigma$ above
the value observed for 1005 GRBs (corrected for the sky exposure bias).
Any GRBs interior to this radius will have to be balanced by GRBs
exterior to this radius.

As our simplest halo model we consider a sphere centered on the Galactic
center containing a uniform density of sources, all of which are observable
by BATSE.
Wasserman (1992) has considered a similar model.
The lowest solid curve of the left panel of Figure~8e shows that
the radius $R_{\rm halo}$ of this sphere must be greater than 194 kpc
to avoid BATSE typically detecting the anisotropy in 1005 GRBs at the
99\% confidence level.

Our next model is based upon a simple functional form for the
Dark Matter Halo used to explain the rotation curves of galaxies including
the Milky Way: density
\begin{equation}
\rho (R) \propto \frac{1}{1+(R/R_{\rm core})^2}
\end{equation}
The $R^{-2}$ dependence yields the observed flat rotation curves and
the core radius prevents an excess contribution at the center
and prevents the function from diverging.
In contrast, the distribution of the luminous halo of our Galaxy is at
least as steep as $R^{-3.5}$ (Gilmore, King \& van der Kruit 1990;
Djorgovski \& Meylan 1994)
which is more anisotropic than the $R^{-2}$ distribution that we use.
Hakkila et al. (1994b) found using 452 GRBs that exponents from
1.5 to 4.0 were acceptable.
Paczy\'{n}ski (1991b) and Mao \& Paczy\'{n}ski (1992b)
used this distribution and concluded that there would
be an observable deficiency of faint sources and an undetectable
dipole moment only if $R_{\rm core} > 2R_\circ = 17 {\rm kpc}$.
Brainerd (1992) found that a somewhat smaller value of $R_{\rm core}$ was
permitted if GRBs are not standard candles.

Figure~8f presents our results for a Dark Matter Halo of standard
candle sources observed to 300 kpc.   This observing depth is near
the maximum permitted without seeing sources from M31 and is near the
best value found by Hakkila et al. (1994a), considering both isotropy and
inhomogeneity.
A smaller observing depth for a fixed number of observed GRBs would
lead to a larger dipole moment.
We find that a core radius of $R_{\rm core}$ = 23 kpc
would be typically found in 1005 locations by BATSE at the 99\% confidence
level.  Hakkila et al. (1994a)
found that the Dark Matter Halo can satisfy both the
isotropy and inhomogeneity requirements of BATSE, however, a larger value
of $R_{\rm core}$ than 23 kpc is required
in order to avoid having too few faint GRBs.

In many halo models the nearest GRBs have a greater gradient and thus
dipole moment than the entire distribution so the anisotropy might be
first detectable in a subset of near GRBs.
This is not the case for the Uniform Spherical Halo
since if we do not view far enough to see its edge, the model becomes
isotropic and homogeneous.
For the Dark Matter Halo model, we find that for small datasets such that
the total dataset limit on $R_{\rm core}$ is $\lesssim R_\circ$, examining the
brightest 1/4 or 1/8 of the GRBs gives a better limit.   For larger datasets
that already require a large value of $R_{\rm core}$, examining a smaller
dataset gives a less strict limit because the fluctuations of the
smaller sample size are more important than the increased dipole moment.
In any case, the brightest sources, e.g., datasets \#24, 25,
31, 32, 38, and 39, are consistent with isotropy.

Smith (1994b) has proposed another geometric halo model, an exponential
halo with source density
$\propto \exp (-R/\overline{r})$.   His example successful model
has $\langle \cos \theta \rangle$ $=0.056$, which deviates by
$2.1\sigma$ from the value observed for 1005 GRBs.    This particular
exponential halo is a fair match for the data.

\subsection{High-Velocity Neutron Star Models}

The Dark Matter Halo model with a large core radius can match the isotropy
and M31 constraints, and also the inhomogeneity constraint
(Hakkila et al. 1994a).
This model is primarily geometric because no candidate GRB source population
has this distribution.
This raises the question of whether the distribution of any
known object can match the observations.
Neutron stars are, of course, the most promising candidates.

Paczy\'{n}ski (1990) and Hartmann, Epstein \& Woosley (1990) calculated
the orbits of neutron stars born in the disk.
They found that if the distribution was viewed deeply enough to detect
a deficiency of faint sources there would be a significant quadrupole
moment $\langle \sin^2 b - \frac{1}{3} \rangle$.
New observations and analyses
since then have demonstrated the existence of a large number
of high velocity neutron stars
(e.g., Lyne \& Lorimer 1994; Frail, Goss, \& Whiteoak 1994).
Lyne \& Lorimer (1994) find the mean neutron star velocity to be
$450\pm90$ km s$^{-1}$ and the velocity distribution to be very broad,
with an root-mean-square width of 525 km s$^{-1}$.
The mean velocity is close to the local speed to escape the
Galaxy, $550\pm100$ km s$^{-1}$ (Fich \& Tremaine 1991).
Many neutron stars are thus either unbound or only weakly bound, implying
the existence of a much larger halo than formerly thought
(Lyne \& Lorimer 1994).
However, these
observations still require a substantial fraction of the neutron stars
to have velocities below the Galactic escape speed.
The low-velocity neutron stars generate a disk population in addition
to the halo formed by the high-velocity neutron stars, so this model
still has difficulty matching the observations.
A variety of solutions have been suggested:
1) a hypothetical population of neutron stars that are born in the halo;
this population either being more numerous than that born in the disk
or more efficient at bursting
(Brainerd 1992;
Eichler \& Silk 1992;
Hartmann 1992;
Fabian \& Podsiadlowski 1993),
2) a delayed turn-on so neutron stars still near the disk are not
seen (Li \& Dermer 1992),
3) some mechanism such as a magnetic-field/recoil velocity correlation
so that only high-velocity neutron stars produce gamma-ray bursts
(Brainerd 1992; Li \& Dermer 1992), and
4) an alignment between the recoil velocity of a neutron star and its
magnetically beamed gamma-ray burst emission (Li et al. 1994).

Eichler \& Silk (1992) propose that neutron stars are born in an extended
halo via the mergers of Population III white dwarfs.    They assume a
Dark Matter Halo distribution with $R_{\rm core}$ = 40 kpc and
estimate $\langle \cos \theta \rangle$ to be 0.05, which
deviates from the observed value by about $1.8\sigma$.
Hartmann (1992) calculated the orbits of neutron stars born
with 400 km s$^{-1}$ recoil velocities in an $R^{-2}$ halo of radius 50 kpc.
An observing depth of 30 kpc matches the intensity observations but
yields $\langle \sin^2 b - \frac{1}{3} \rangle$ $= -0.05$, $5.0\sigma$
off from the current observations.
Fabian \& Podsiadlowski (1993) propose that gamma-ray burst sources
are ejected from the Magellanic Clouds, producing an extended halo nearly
homogeneous to 50 kpc.
Their example model has a dipole moment to the Magellanic Clouds of
0.038, which deviates by $1.9\sigma$ from observed moment (Table~8).
The agreement of these models with the data ranges from satisfactory
to unacceptable.

Li \& Dermer (1992) proposed that gamma-ray bursts originate
after a time delay from neutron stars born in the disk with velocities
of 1000 km s$^{-1}$.
The time delay allows many of the sources to leave the disk before they
become active, thereby reducing the disk contribution.
Their model has
$\langle \cos \theta \rangle$ = 0.048, which is $1.7\sigma$ off from
the current observed value.

The Halo Beaming Model of Li et al. (1994) assumes that 1) only high-velocity
neutron stars emit GRBs, 2) these objects are
born with their magnetic axis aligned with their recoil velocity, and
3) the gamma-ray burst emission is beamed in a $20^\circ$ half-angle beam along
the magnetic axis.
This beaming suppresses the observation of sources in the plane relative
to those far above the plane.
Li et al. (1994) determine the consequences of this model by following
the orbits of  neutron stars born in the disk with recoil velocities
of 1000 km s$^{-1}$.
This model still predicts a concentration in the plane, and, uniquely,
a quadrupole concentration towards the Galactic center and anti-center; both
concentrations should be most prominent for the nearest GRBs.
The authors
show, as a function of peak flux on the 1024 ms timescale,
the model values of $\langle \sin^2 b - \frac{1}{3} \rangle$ and
$\langle \cos^2 \theta - \frac{1}{3} \rangle$
for sets of all GRBs brighter than the peak flux value.
The brightest 1/8 of the first 1005 GRBs deviate from the model values
by $2.0\sigma$ and $3.1\sigma$ (Table~8).
This particular form of the Halo Beaming Model seems unlikely.

The models of Podsiadlowski et al. (1995) use several methods to reduce the
anisotropy of neutron stars ejected from the plane.
They follow the orbits of
neutron stars in a non-spherical halo potential, which
randomizes the orbits, thereby producing greater isotropy.
In order to avoid a disk signal, they
assume that only high-velocity neutron stars burst and
that the sources have a delayed turn-on and a turn-off time.
The latter requirement prevents the build of of a disk population formed
by the capture of neutron
stars that received their birth kick opposite to the Galactic rotation.
Their example models (their Figures~5a and 5b) are based upon:
1) standard candle sources observed to 340 kpc,
2) neutron stars born with recoil velocities between 600 and 700 km s$^{-1}$,
3) a turn-on delay of $10^7$ years,
4) turn-off times of $10^9$ years for model~5a and $10^{10}$ years for
model~5b.
The values of
$\langle \cos \theta \rangle$ and
$\langle \sin^2 b - \frac{1}{3} \rangle$ of their model~5b are $2.0\sigma$ and
$2.2\sigma$ off from those observed for 1005 GRBs, demonstrating that
a source lifetime of $10^{10}$ years is probably too long.
The moments of the $10^9$ year lifetime model are in good
agreement with the data, deviating by $1.4\sigma$ and $1.7\sigma$.

\subsection{Two-Component Models}

It has frequently been suggested that GRBs might originate from two
populations in various combinations of disk, halo and cosmological distances
(e.g.,
De Jager \& Schaefer 1992;
Lingenfelter \& Higdon 1992;
Smith \& Lamb 1993;
Higdon \& Lingenfelter 1994;
Katz 1994).
Since sufficiently extended halo models agree with the observations we
expect a small admixture of a disk component with either a halo or the
isotropic distribution to also be satisfactory.
As an example we consider a model with fraction
$f_{\rm aniso}$ of the locations
drawn from the Watson distribution with $\kappa=-1$ and the
remainder isotropic.
The Watson distribution with $\kappa=-1$ has one-half of the
probability within $b_{\frac{1}{2}} = 23.1^\circ$ of the plane and
$\langle \sin^2 b - \frac{1}{3} \rangle$ = $-0.0796$, an $8.4\sigma$ detection
for 1000 GRBs; an example simulation is shown in Figure~1c.   From
Figure~8g we see that BATSE would typically detect at the 99\%
confidence level a 30\% contribution from this disk model in
a sample of 1000 locations.
Of course, the isotropy constraints would allow a greater fraction
of a less concentrated disk component.
However, the contribution of an almost isotropic component is limited by
its near homogeneity.

Lingenfelter \& Higdon (1992) consider a model with two Galactic components
arising from a single source population.   The low-luminosity emissions are
seen to a nominal distance of 300 pc and the high-luminosity to 100 kpc.
The source population is based
upon the orbital calculations of Paczy\'{n}ski (1990) for
Galactic neutron stars born in the disk.
The sources seen at the great distances contribute most of
the anisotropy---the low-luminosity sources are
virtually isotropic and homogenous.
This 1992 model now clearly disagrees with the data,
differing from the observations for
1005 GRBs at the
$3.5\sigma$ level for $\langle \cos \theta \rangle$ and at the
$6.1\sigma$ level for $\langle \sin^2 b - \frac{1}{3} \rangle$.
Their revised two-component model (Higdon \& Lingenfelter 1994) is formed from
two geometric models: an exponential disk and a
Dark Matter Halo model with a cutoff at 300 kpc.
Since at most 25\% of the GRBs are from the disk,
the dominant anisotropies of their models are those of the halo portions,
the dipole moment $\langle \cos \theta \rangle$.
The comparisons are given in Table~8.   The core radii of 7.5 and
15 kpc, which are ``inferred from constraints on Galactic structure'',
are in very poor agreement with the observations.
Even the model with a halo component with a core radius of 30 kpc does not
agree well with the observations.    This last model has a disk component
of only 8\%.

Smith \& Lamb (1993) consider several other disk plus halo models.
Their disk model is an exponential sampled to 2/3 of its scale height,
resulting in a quadrupole moment $\langle \sin^2 b - \frac{1}{3} \rangle$
$= -0.0401$, which is half the moment of our example two-component
model (above, Figures 1c, 8g).
One of their models has 20\% of the sources from the disk, while the
remainder are from a Dark Matter Halo with core radius 22.5 kpc observed to
135 kpc.   Again, this is predominantly a halo model, and it is currently
$2.2\sigma$ away from BATSE's measurement of
$\langle \cos \theta \rangle$.
Their other model has 67\% of the GRBs originating in the disk and
the remainder from a Gaussian shell halo model.    The halo model is a
Gaussian distribution with $\sigma$ = 38 kpc and its highest source density
on a shell 25 kpc from the Galactic center.   Since this model is
dominated by its disk component, the crucial statistic is
$\langle \sin^2 b - \frac{1}{3} \rangle$ = $-0.027$,
which is $2.6\sigma$ away from
the value measured by BATSE for 1005 GRBs.

In this paper we are emphasizing the isotropy characteristics of two-component
models.   It is also necessary for models to match the observed
peak flux distribution, $\log{N}-\log{P}$.
A particular requirement for two-component
models to agree with BATSE's observations is that the
slope of the logarithmic cumulative
peak flux distribution be concave down
(Paczy\'{n}ski 1992; Smith \& Lamb 1993; Hakkila et al. 1994a; Harding 1994).

\subsection{Heliocentric Models}

No heliocentric pattern is evident (\S4).
The lack of a dipole moment towards the Sun is a weak constraint:
a spherical shell at the 19 AU distance of Uranus would produce a dipole
moment of $\langle \cos \phi \rangle$ = 0.035, which would be a
$1.9\sigma$ detection for 1005 GRBs
(adapting eq. 1 from galactocentric to heliocentric shells).
The above limit applies to the ``typical'' gamma-ray burst; testing
the plane-wave nature of the wavefront of bright and presumably close
bursts yields lower-limits of $\approx 40$ AU (Conners et al. 1993).

The models that have been suggested for a heliocentric origin
involve comet-comet or comet-black hole collisions in the Oort cloud
(Bickert \& Greiner 1993; Katz 1993; White 1993, Luchkov 1994).
One can therefore constrain these models based upon our knowledge of
the cometary cloud surrounding the Sun: models
of the distribution of the heliocentric cometary cloud are based
upon a solar system origin, planetary and stellar perturbations and the
tidal force of the Galaxy (Weissman 1990; Clarke, Blaes \& Tremaine 1994).
The Galactic tidal force has had negligible effect on the inner region of the
cloud, which should thus have a strong ecliptic plane concentration
reflecting its solar system origin (Clarke et al. 1994).
The lack of an ecliptic plane concentration according to
$\langle \sin^2 \beta - \frac{1}{3} \rangle$
is thus a useful constraint.
In contrast, the distribution of the intermediate Oort cloud is
dominated by Galactic tidal forces---Clarke et al. (1994) estimate
$\langle \sin^2 b - \frac{1}{3} \rangle$ = +0.12, which is quite
inconsistent with the observed GRB locations.
While the distant portion of the cloud is probably more isotropic,
producing GRBs there but not in the inner portion is very implausible
since the cometary density is lower and the comets are more distant
(Clarke et al. 1994).

A severe problem for Oort cloud models of GRBs is
the homogeneity of the brightest GRBs
(Maoz 1993b; Clarke et al. 1994; Horack et al. 1994), which
has been well-observed by instruments with greater exposure than
BATSE (e.g., Fenimore et al. 1993).
The physics of the models pose even greater problems
(Clarke et al. 1994).

\section {Discussion}

If gamma-ray bursts have a large-scale anisotropy, it is remarkably
small, especially compared to any observed Galactic component.
The most extended, observed component of the Galaxy is the halo,
as traced by stars or globular clusters.
The globular clusters are distributed in galactocentric radius $R$
as $R^{-3.5}$ (Zinn 1985; Gilmore, King \& van der Kruit 1990;
Djorgovski \& Meylan 1994).
This contrasts with the more isotropic
$R^{-2}$ we used in the Dark Matter Halo form (eq. 2).
Using a slightly different halo form than
ours, Djorgovski \& Meylan (1994) find a core radius of at most 2 kpc
for the globular clusters, which contrasts with the $2\sigma$ lower limit
for the core radius of the GRBs of 11 kpc.
Zinn (1985) has shown that the globular clusters consist of two
populations, a thick disk and a halo, and that population membership is
reliably determined by cluster metallicity.
The 93 globular clusters identified by Zinn (1985) as halo members
have $\langle \cos \theta \rangle$ = 0.617 and
$\langle \sin^2 - \frac{1}{3} \rangle$ = $-$0.111, which are, respectively,
$10.3\sigma$ and $3.6\sigma$ deviations from isotropy!
Selection effects against globular clusters near the Galactic plane or
center make the observed distribution more isotropic than the true
distribution (Zinn 1985, Djorgovski \& Meylan 1994).
These moments contrast sharply with those of the first 1005 GRBs observed by
BATSE: the $2\sigma$ limits corrected for sky exposure bias are
$\langle \cos \theta \rangle < 0.053$ and
$\langle \sin^2 - \frac{1}{3} \rangle > -0.022$.

While geometric Galactic
models for the distribution of GRBs can match the observed
isotropy, they are different from any known Galactic distribution.
Although there is observational and theoretical
support for an exponential disk, this is not the case for an exponential halo
(Gilmore, King \& van der Kruit 1990).
We know of no Galactic component with its peak density
25 kpc from the Galactic center, as is assumed in the Gaussian shell
halo model.
The maximum core radius for the Dark Matter Halo component of the Galaxy models
of Caldwell \& Ostriker (1981) is about 9 kpc.
While the $2\sigma$
upper limit for $\langle \cos \theta \rangle$ implies a core radius
of at least 11 kpc, taking account of the GRB peak flux distribution
demands a much larger core radius (Hakkila et al. 1994a).
The model of Higdon \& Lingenfelter (1994) with a core radius of 15 kpc
has a value of $\langle \cos \theta \rangle$ which is $3.1\sigma$ in
excess of the value observed for 1005 GRBs (Table~8).
While previously
core radii consistent with the Dark Matter Halo form as used to
explain the rotation curve of the Milky Way were consistent with BATSE's GRB
observations, this is no longer true.
Our geometric Dark Matter Halo assumed spherical symmetry, whereas
Dark Matter Halos are actually expected to be quite flat
(Dubinski \& Carlberg 1991).

Possible physical bases for the Dark Matter Halo distribution
for GRBs are that the $R^{-2}$ dependence is similar to that of
an isothermal halo
(Gilmore, King \& van der Kruit 1990) and to the density of very-high
velocity objects ejected from the Galaxy.
Many recent models have been based upon the assumption that the sources
of GRBs are neutron stars born with velocities comparable to or
greater than the escape speed from the Galaxy.
The mechanism postulated
to cause only high-velocity neutron stars to burst
must be very effective.
The fraction of observed GRBs originating from
low-velocity neutron stars, which are confined to a thick disk, must
be small (\S5.7).
The fraction of GRB sources which have low velocities must be yet
smaller since many of the high-velocity neutron stars are unbound and
escape the observable region, which must have a radius
$\lesssim$ 350 kpc because of the M31 constraint.
Even the assumption that only high-velocity neutron stars produce
GRBs is insufficient to match the data---all of the models have at least
one additional assumption:
1) a delayed turn-on (Li \& Dermer 1992; Podsiadlowski et al. 1995),
2) a turn-off time (Podsiadlowski et al. 1995),
3) a population of high-velocity neutron stars born in the halo which
dominate neutron stars born in the disk as GRB sources
(Eichler \& Silk 1992; Hartmann 1992),
4) a preferential production of GRB sources in
the Magellanic Clouds (Fabian \& Podsiadlowski 1993), and
5) an alignment between the gamma-ray emission and the recoil
velocity of the neutron star (Li et al. 1994).

Since a very-extended halo is consistent with the data,
it is possible to have two-component models with one or two Galactic
components.
The observations currently require that at most a small fraction of the bursts
originate from a disk component (\S5.7).
A common motivation for two population models is allow GRBs with
cyclotron lines to originate from disk neutron stars in the
solar neighborhood.
Models of photon emission from magnetic neutron stars
were developed when it was generally believed that the
distance to GRBs was of order 100 pc.   It seems unlikely that these models
will work at halo distances which are $\gtrsim$ 100 times greater.
Two population models generally have a ratio of the distances of the
populations of at least 100, implying a ratio of luminosities of order
$10^4$ or more.
It seems implausible that phenomena with such greatly different
luminosities would not be observationally separable into two classes.
The observed bimodality in GRB durations
(Mazets et al. 1981; Hurley 1991; Klebesadel 1992; Dezalay et al. 1992;
Kouveliotou et al. 1993b)
is the best indication for two possible classes, but both the short and
long GRBs are consistent with isotropy (datasets \#14 and \#15)
and inconsistent with homogeneity (Kouveliotou et al. 1993b).
Invoking two populations to explain the GRB position and intensity
distributions
would seem to double the difficulty of
explaining the physics of gamma-ray bursts.

Several of the Galactic models have moments in acceptable agreement
with the data---the best
has a moment $1.7\sigma$ in excess of the observations (Table~8).
However, most of the Galactic models created in response to BATSE's
observations
have moments that deviate by two or more $\sigma$ from the data.
While if GRBs are Galactic, only the true model need predict moments
consist with the observations, the failure of the majority of the post-BATSE
models demonstrates the difficulty of matching Galactic models to
BATSE's observations.
Each model has been tested by comparing the observations
with the model characteristics for at most
one dipole and one quadrupole moment.
Further tests of the models are possible, such as comparing
their radial distributions with the observed $\log{N}-\log{P}$
distribution.

The simultaneous observations by BATSE
of the isotropy and the strong inhomogeneity
of GRBs disprove the previous paradigm
of a Galactic disk origin.
No known Galactic component, including the Dark Matter halo, is sufficiently
isotropic to match the observational constraints.
Consequently, current models either postulate a new Galactic component or
extreme energy requirements for sources at cosmological distances.
If GRBs are Galactic, the lack of excess GRBs from the Galactic Center
and M31 constrain the distance of the typical GRB observed by BATSE
to between 100 and 300 kpc (\S5.5).
In contrast to the difficulty that the isotropy observations pose for Galactic
models, isotropy is a natural consequence of an origin at cosmological
distances.
The observed large-scale isotropy
strongly favors but does not require that gamma-ray
bursts originate at cosmological distances.
The continued observation of gamma-ray bursts by BATSE promises to either
detect an anisotropy or to further tighten the constraints on Galactic models.

\acknowledgments
M. S. Briggs acknowledges support provided by NASA through the
Compton GRO Fellowship Program, grant GRO/PFP-91-06.
D. Hartmann acknowledges support from NASA grant NAG-51578.
We thank J. J. Brainerd, D. Lamb and the referees,
J. Higdon and P. Podsiadlowski, for their useful suggestions,
and the BATSE operations team for their efforts.

\clearpage

\setcounter{page}{37}
\begin{deluxetable}{ccccc}
\tablecolumns{5}
\tablenum{2}
\tablewidth{0pt}
\tablecaption{ANGULAR ERRORS: DATASET \#6 AND SIMULATIONS}
\tablehead{
\multicolumn{3}{c}{2B less MAXBC}  &   \multicolumn{2}{c}{Simulations}   \\
\colhead{Fraction}  &   \colhead{$\sigma_{\rm stat}$}    &
\colhead{$\sigma_{\rm tot}$}    &   \colhead{Fraction}  &
\colhead{$\Sigma$}    \\
\colhead{}   &    \colhead{(degrees)}    &    \colhead{(degrees)}   &
\colhead{}   &    \colhead{(degrees)}   }
\startdata
50\%    &    0.1---3.7    &    4.0---5.4    &    45\%    &    15.   \\
25\%    &    3.7---7.4    &    5.4---8.4    &   22.5\%   &    20.   \\
18.8\%  &    7.4---13.8   &    8.4---14.4   &   16.875\% &    30.   \\
6.2\%   &   13.9---29.9   &   14.5---30.1   &   15.625\% &    60.   \\
\enddata
\end{deluxetable}

\clearpage

\begin{deluxetable}{ccc}
\tablewidth{255pc}
\tablecolumns{3}
\tablenum{3}
\tablewidth{0pt}
\tablecaption{EXPECTED VALUES OF THE STATISTICS:
CORRECTED FOR BATSE'S SKY EXPOSURE}
\tablehead{
\colhead{Statistic}    &  \colhead{Mean\tablenotemark{a,b}} &
\colhead{$\sigma$}  }
\startdata
$\langle \cos \theta \rangle$                  &  $-$0.013  &
0.99$\sqrt{ \frac{1}{3N_B} }$  \\
$\langle \sin^{2} b - \frac{1}{3} \rangle$     &  $-$0.005  &
0.99$\sqrt{ \frac{4}{45N_B} }$  \\
${\cal W}$            & $3+0.0020N_B$  & \nodata \tablenotemark{c}  \\
${\cal B}$            & $5+0.0077N_B$  & \nodata \tablenotemark{c}  \\
$\langle \sin \delta \rangle$                  &  0.026 &
1.04$\sqrt{ \frac{1}{3N_B} }$  \\
$\langle \sin^{2} \delta - \frac{1}{3} \rangle$ &  0.026 &
1.03$\sqrt{ \frac{4}{45N_B} }$  \\
$\langle \cos \phi \rangle$  &
$0.010 \cos \left[ \frac{2\pi({\rm T}-9160)}{365.25} \right]$  &
0.98 $ \sqrt{ \frac{1}{3N_B} }$  \\
$\langle \sin^2 \beta - \frac{1}{3} \rangle$  &   0.020  &
$1.02 \sqrt{ \frac{4}{45N_B} }$  \\
\enddata
\tablenotetext{a}{These means are valid for time intervals long enough
to average over the CGRO precession period of about 50 days.}
\tablenotetext{b}{$T$ is the time in Truncated Julian Day Number (TJD),
which is the Julian Day Number minus 2440000.5}
\tablenotetext{c}{see Figure~4.}
\end{deluxetable}

\clearpage

\setcounter{page}{40}
\begin{deluxetable}{rlrcccccc}
\tablecolumns{9}
\tablenum{5}
\tablewidth{0pt}
\tablecaption{RESULTS: EARTH COORDINATE SYSTEM TESTS}
\tablehead{
\multicolumn{3}{c}{Dataset}   &
\multicolumn{3}{c}{$\langle \sin \delta \rangle$}  &
\multicolumn{3}{c}{$\langle \sin^{2} \delta - \frac{1}{3} \rangle$}  \\
\colhead{\#}  &   \colhead{Name}    &  \colhead{Size}  &
\colhead{Value}  &   \colhead{Dev.\tablenotemark{a}}   &
\colhead{Raw Dev.\tablenotemark{b}}  &   \colhead{Value}   &
\colhead{Dev.\tablenotemark{a}}    &   \colhead{Raw Dev.\tablenotemark{b}}   }
\startdata
2   & 1B revised  &  262  &  0.015  &  $-$0.3    &      +0.4      &
                             0.000  &  $-$1.3    &  $\ $ 0.0   \\
3   & 2B          &  585  &  0.003  &  $-$0.9    &      +0.1      &
                             0.024  &  $-$0.1    &      +2.0  \\
1   & first 1005  & 1005  &  0.016  &  $-$0.5    &      +0.9      &
                             0.019  &  $-$0.7    &      +2.0   \\
\enddata
\tablenotetext{a}{
Deviation, in $\sigma$, of the observed value from the value calculated from
BATSE's sky exposure map assuming isotropy.}
\tablenotetext{b}{
Deviation, in $\sigma$, of the observed value from the value expected for
isotropy ignoring BATSE's sky exposure.}
\end{deluxetable}

\end{document}